# SolarEV City Concept for Paris: A promising idea?[1]

Paul Deroubaix[1,2], Takuro Kobashi[3,*], Léna Gurriaran[2], Fouzi Benkhelifa[4], Philippe Ciais[2], Katsumasa Tanaka[2,5*]

[1] École Polytechnique, Palaiseau, France

[2] Laboratoire des Sciences du Climat et de l'Environnement, IPSL, CEA-CNRS-UVSQ, Université Paris-Saclay, Gif-sur-Yvette, France

[3] Graduate School of Environmental Studies, Tohoku University, Sendai, Japan.

[4] NEXQT – Liberté Living Lab

[5] Earth System Division, National Institute for Environmental Studies, Tsukuba, Japan

\* Corresponding authors
Emails: takuro.kobashi.e5@tohoku.ac.jp / katsumasa.tanaka@lsce.ipsl.fr

**Abstract.** Urban decarbonization is one of the pillars for strategies to achieve carbon neutrality around the world. However, the current speed of urban decarbonization is insufficient to keep pace with efforts to achieve this goal. Rooftop photovoltaics (PVs) integrated with electric vehicles (EVs) as battery is a promising technology capable to supply $CO_2$-free, affordable, and dispatchable electricity in urban environments ("SolarEV City Concept"). Here, we evaluated Paris, France for the decarbonization potentials of rooftop "PV + EV" in comparison to the surrounding suburban area Ile-de-France and Kyoto, Japan. We assessed various scenarios by calculating the energy sufficiency, self-consumption, self-sufficiency, cost savings, and $CO_2$ emission reduction of the PV + EV system or PV only system. We found that above a certain roof coverage, that is, 50-60% of the total roof area for Paris or 20-30% for Ile-de-France, PV electricity production regularly exceeds the demand, resulting in a lower self-consumption. Above that roof coverage, feed-in-tariffs (FIT) or storage are needed to further exploit the potential of the PV + EV system. The combination of EVs with PVs by vehicle-to-home (V2H) or vehicle-to-building (V2B) systems at the city or region level was found to be more effective in Ile-de-France than in Paris suggesting that SolarEV City is more effective for geographically larger area including Paris. If implemented at a significant scale, they can add substantial values to rooftop PV economics and keep a high self-consumption and self-sufficiency, which also allows bypassing the classical battery storage that is too expensive to be profitable. Furthermore, the systems potentially allow rapid $CO_2$ emissions reduction; however, with already low−carbon electricity of France by nuclear power, $CO_2$ abatement (0.020 $kg_{CO2}$/kWh reduction from 0.063 $kg_{CO2}$/kWh) by PV + EV system can be limited, in comparison to that (0.270 $kg_{CO2}$/kWh reduction from 0.352 $kg_{CO2}$/kWh) of Kyoto, also because of the Paris's low insolation and high demands in higher latitude's winter. While the SolarEV City Concept can help Paris to move one step closer to the carbon neutrality goal, there are also implementation challenges for installing PVs in Paris.

---

[1] The short version of the paper was presented at ICAE2022, Bochum, Germany, Aug 8-11, 2022. This paper is a substantial extension of the short version of the conference paper.

**Highlights**
- From roof coverage of 50-60% PV production regularly exceeds demand in Paris.
- PV + EV system is more effective in Ile-de-France than Paris for larger roof area.
- PV + EV system is less effective in Paris than in Kyoto for climate.

**Introduction**

Many state governments including the US, China, India, and Japan have announced so-called "net zero" emission targets by mid-century to steer the world towards an emission pathway consistent with the long-term temperature goal of the Paris Agreement [1,2]. In July 2021, the EU established a legal framework to achieve net zero greenhouse gas (GHG) emissions by 2050 [3], with individual states setting varied target years (e.g. 2045 for Germany; 2050 for France). Net zero targets have also been put forward by regional and local municipalities including Paris and Kyoto. Decarbonization of urban areas is important because $CO_2$ emissions from cities around the world account for 71%-76% of the global $CO_2$ emissions [4], and because first-hand decarbonization measures are often performed through local governments [5]. Increasing zero carbon energy technologies (e.g., renewables) and improving energy efficiency, as well as electrifying the transport and heat system, are major pillars of decarbonization [6,7]. The SolarEV City concept [8], a novel concept that makes use of synergies between rooftop photovoltaics (PVs) and electric vehicles (Evs) as battery (PV + EV), supports such pillars. When both technologies are combined, EVs can not only eliminate $CO_2$ emission from gasoline/diesel combustion through PVs, but also serve as electricity storage to address intermittency with rooftop PVs [9,10].

The first study for the "PV + EV" in a city scale was conducted for Kyoto City, Japan (hereafter, "Kyoto"), which demonstrated that by using 70% of the rooftop area of Kyoto for PVs and converting all passenger vehicles to EVs, Kyoto's $CO_2$ emission from electricity generation and ICEs (internal combustion engine vehicles) can be reduced by 60-74% with 22-37% energy cost reduction in 2030 [9]. Then, the analyses were extended to nine Japanese urban areas: Koriyama City, Sendai City, Okayama City, Kyoto City, Kawasaki City, Hiroshima City, Sapporo City, Niigata City, and Tokyo special districts [8]. It was found that the effectiveness of "PV + EV" systems for urban decarbonization is highly variable, depending on the level of urbanization of the cities. For example, Kawasaki and Tokyo are highly urbanized areas with a small rooftop area and the number of vehicles per capita, resulting in relatively limited $CO_2$ emission reduction (54-58%) by the PV + EV systems with energy cost saving of 23-26%. On the other hand, regional cities such as Niigata and Okayama have a larger rooftop area and the number of vehicles per capita, which can yield larger $CO_2$ emission reduction (up to 95%) by the systems with higher energy cost saving of up to 34%.

The effectiveness of the "PV + EV" systems also depend on the country specific factors such as electricity tariff, climate, and urban structures. Chang et al. analyzed five South Korean cities, which indicated "PV + EV" systems can reduce $CO_2$ emissions up to 86% with cost saving of 51% [11]. South Korean Cities have generally higher buildings than those in Japanese cities, which reduces the benefits of "PV + EV" systems than those of Japanese cities [11]. Liu et al. analyzed Shenzhen, China, which is a highly urbanized new city, and found that $CO_2$ emissions can be reduced by 42% and cost saving by 21% through the "PV + EV" system [12]. Dewi et al. evaluated the potentials of the "PV + EV" system for Jakarta, the capital of Indonesia, and found $CO_2$ emission reductions by 76-77%, accompanied by energy cost savings by 33-34% [10].

The study by Arowolo and Perez analyzed Paris, Lyon, and Marseille, France for the effects of "PV + EV" systems [13]. For a half of rooftop available space installed with PVs and a half of passenger vehicles replaced with EVs, they found 20-42% of electricity demand reduction, 43-48% of $CO_2$ emission reduction, and a payback period of 2-3 years by 2030. Although Arowolo and Perez analyzed Paris for the PV + EV system as we did, they did not account for the electricity supply to the city from rooftop PV generation by considering hourly supply-demand balance (self-sufficiency). It is crucial to consider the benefit of electricity supply from PVs with supply-demand balance when evaluating the potential of the system for Paris due to its high-latitude location with strong seasonal variation of solar insolation. Paris can be seen as an iconic city of climate change as it is the birthplace of the Paris Agreement, so is Kyoto for the Kyoto Protocol. Our analysis of Paris and Kyoto may thus be useful for raising awareness of the SolarEV City Concept.

Under the background above, this paper will strive to answer the following questions:
1. How can the SolarEV City Concept (rooftop PV system combined with EVs as battery) contribute to the decarbonization of the City of Paris (high latitude city)?
2. What are the factors that can affect the potential of "PV + EV" for Paris in comparison to those of Ile-de-France and Kyoto?
3. What would be barriers to implement the SolarEV City Concept in Paris?

This paper is organized as follows. Section 2 describes the methods and materials of this study, including the city of Paris, techno-economic analysis, and scenarios. Section 3 illustrates the results of the analyses of Paris in comparison with those of other regions. Section 4 discusses the implications of the results and barriers to implement such technologies in a city scale. Section 5 concludes the paper.

**2. Methods and materials**
2.1. "Paris" and surrounding region, "Ile-de-France"

We studied two areas: Paris "intramuros", i.e. the city without its suburbs, and Ile-de-France, which is the region around Paris. Paris is a highly urbanized area with population of over 2.1 million and located in high latitude (48.9° N, 2.4° W) (Fig. 1; Table 1). To fully elucidate the potential of rooftop PVs integrated with EVs for Paris, it is useful to consider the surrounding region of Ile-de-France as potential solar power suppliers (Fig. 1). Ile-de-France (including Paris) covers 18.9% of the population of France (i.e., mainland) [14], which is the largest metropolitan area within the EU. While Paris is an urban area with a high population density and energy intensity, Ile-de-France is mostly a suburban area with a lower population density and energy intensity (Fig. 1). Public transport is well developed in Paris. In Ile-de-France, on the other hand, residents more frequently use and own passenger vehicles. These differences are apparent in the number of registered vehicles per capita and average driving distances (Table 2). Data for Kyoto is also listed for comparison.

The French government will ban the sales of gasoline and diesel vehicles from 2040. Paris government will start an equivalent ban 10 years earlier, and Ile-de-France government set a ban for diesel vehicle sale by 2030. Since 2018, a territorial climate-air-energy plan has been adopted in France, mandating inter-municipalities with more than 20,000 inhabitants to establish a global warming mitigation strategy based on renewable energy and energy consumption control. In this context, the city of Paris has announced that by 2050 it will achieve an energy mix of 100% renewables, of which at least 20% will be supplied locally [15].

Table 1. General statistics of Paris, Ile-de-France, and Kyoto. Data for Kyoto are from [16].

|  | Paris | Ile-de-France | Kyoto |
| --- | --- | --- | --- |
| Population (million) | 2.18 | 12.2 | 1.47 |
| Area ($km^2$) | 105 | 12,011 | 827 |
| Density (1000ppl/ $km^2$) | 20.6 | 1.0 | 1.8 |
| Roof area ($km^2$) | 31 | 402 | 52 |
| Roof area per capita ($m^2$) | 14 | 33 | 13 |
| Annual Electricity consumption per capita (kWh) | 6031 | 6277 | 5678 |

Table 2. The number of cars, the state of car use, and the average travel distance in Paris, Ile-de-France, and Kyoto. Data for Kyoto are from [16].

|  | Paris | Ile-de-France | Kyoto |
| --- | --- | --- | --- |
| Number of diesel cars (thousand) | 250 | 2,802 | - |
| Number of gasoline cars (thousand) | 334 | 2,525 | - |
| Total number of cars (thousand) | 585 | 5,327 | 485 |
| Number of cars per capita | 0.27 | 0.44 | 0.33 |
| Proportion of cars used during weekdays | 0.35 | 0.63 | - |
| Average distance of driving cars (km per car per day) | 21.8 | 26.3 | 8.0 |

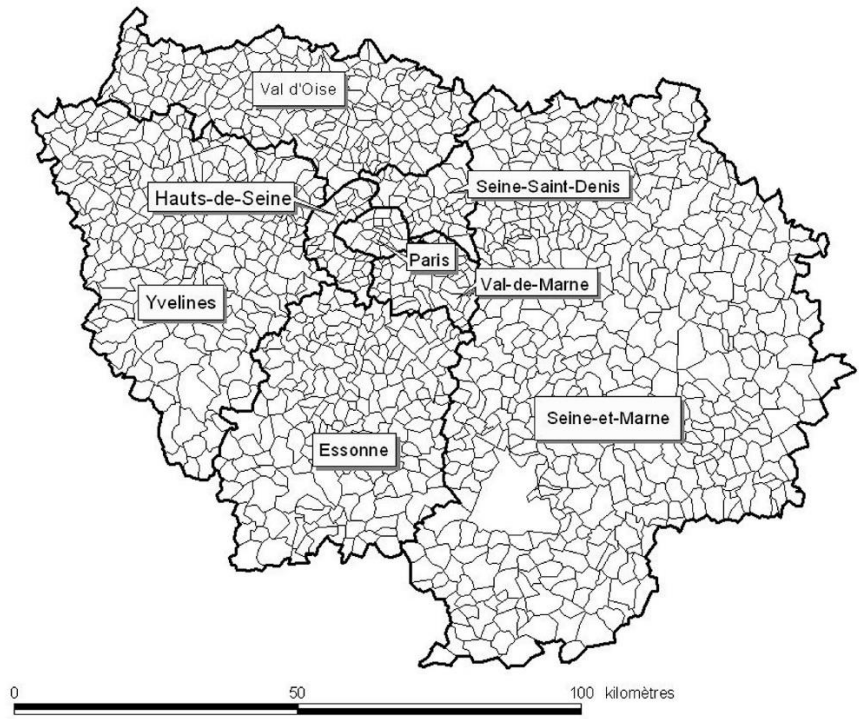

Figure 1. Map of Ile-de-France including the city of Paris ("Paris Intramuros").

2.2. Techno-economic analysis

The viability of variable renewable energy (VRE) projects depends on various factors such as climate, electricity tariff, costs of renewable energy technologies, electricity demand, supply-demand matching, degradation of technologies, feed-in-tariff (FIT), discount rate, urban structures, etc. [17–20]. A techno-economic analysis is a useful tool to evaluate all these factors and can provide information on whether to invest in projects under consideration [19,21]. We employed a methodology developed in the previous studies [16,22] that has been applied to investigate the future potential of the SolarEV City concept elsewhere. The methodological description here is kept concise at the level required for presenting our results.

Analyses were performed with a publicly available energy-economic software, SAM (System Advisor Model; version 2020.11.29 Revision 2), developed by the U.S. Department of Energy's National Renewable Energy Laboratory (NREL) [23]. SAM is designed to evaluate the viability of various types of renewable energy projects [24]. The model runs with a given set of parameters or with a range of values for a subset of parameters iteratively. SAM can analyze the system at the scale of the city (or the region) and considers the surplus and deficit only at this level. The model does not account for the surplus and deficit at the household level, which may well occur without any surplus or deficit at the regional level. Thus, the basic assumption is that rooftop PV generated electricity was consumed in a city in the most-efficient way within the framework we considered. The SAM files used for the analyses are available at Mendeley Data [25].

We calculated the cost savings in terms of net present value of cashflow for a project period of 25 years [16], with a nominal discount rate of 2.5%. The potential changes in the costs of centralized electricity production, in particular the cost of maintaining the current means of production, are not considered. The costs of replacing existing gasoline and diesel vehicles with EVs and the resulting saving of fuel costs are treated as the background in our scenarios and are thus not considered in our cost calculations unless otherwise noted. The net present value is expressed in 2019 euros or 2030 euros, depending on the 1st year of the project. In addition to cost savings, we calculated energy sufficiency, self-sufficiency, self-consumption, and $CO_2$ emission reduction as decarbonization indicators [16]. Energy sufficiency is defined as the total amount of electricity produced by PV in one year divided by the annual electricity demand. Self-sufficiency is the actual electricity coming from PV consumed in a year divided by the annual demand. Self-consumption is the proportion of PV electricity that is consumed within the region considered [16]. More precisely, these indicators [26] can be expressed as:

$$Cost\ saving\ (\%) = \left\{1 - \frac{\frac{NPV}{N}}{AnnualEnergyCost_{Base}}\right\} \times 100$$

where *NPV* is net present value, *N* is project period, and *AnnualEnergyCost$_{Base}$* is annual energy expense of a base scenario (i.e., annual electricity from power company and gasoline cost).

Energy sufficiency = ($E_{PV\text{-}load}$ + $E_{battery\text{-}load}$ + $E_{PV\text{-}gid}$) /($E_{load}$)
Self-sufficiency = ($E_{PV\text{-}load}$ + $E_{battery\text{-}load}$)/ ($E_{load}$)
Self-consumption = ($E_{PV\text{-}load}$ + $E_{battery\text{-}load}$)/ ($E_{PV\text{-}load}$ + $E_{battery\text{-}load}$ + $E_{PV\text{-}grid}$)

where
$E_{load}$ = Total electricity load (kWh·yr$^{-1}$) in the city.
$E_{PV\text{-}load}$ = Electricity (kWh·yr$^{-1}$) supplied from PV directly to load in the city.
$E_{battery\text{-}load}$ = Electricity (kWh·yr$^{-1}$) supplied from battery to load in the city.
$E_{PV\text{-}grid}$ = Electricity (kWh·yr$^{-1}$) exported from PV to grid out of the city.

The outputs from SAM were used to calculate these indices. We calculated $CO_2$ emission reduction from driving and consumption of electricity to assess the environmental benefits. Kyoto was similarly analyzed [16].

2.3. Scenarios

We focused on two illustrative scenarios: "PV only" and "PV+EV". "PV only" assumes a penetration of building rooftop PV. The "PV+EV" scenario assumes a penetration of residential rooftop PV combined with electric vehicles with bi-directional charging (e.g., V2H). In the "PV+EV" scenario, all vehicles are assumed to be electric and connected to vehicle-to-home (V2H) or vehicle-to-building (V2B) systems (bi-directional EV chargers), in which electric vehicles can be used to charge electricity from the rooftop PV and to discharge to the building.

We studied these two scenarios for two areas: Paris "intramuros", i.e. the city without its suburbs, and Ile-de-France, which is the region around Paris, comparing them also with Kyoto [22]. We considered two periods: 2019 (for "PV only") and 2030 (for "PV only" and "PV + EV"). The choice of area affected technical characteristics in the analysis, such as the roof area, the number of cars or the electricity demand. In our analyses, the choice of period only affected the prices, with projections used for the 2030 scenarios, to illustrate the effects of cost decline. For each scenario, we tested a different set of hypotheses on the roof coverage, different technologies ("PV only" or "PV + EV"), cost of technologies ("2019" or "2030"), and feed-in-tariffs ("with" and "without"). As for the techno-economic analysis, we compared the scenarios with a "base" one with the uses of gasoline/diesel engine vehicles and grid electricity [22].

2.4. Meteorological data

We used hourly data for the following parameters: diffuse horizontal irradiance (DHI; $W/m^2$), direct normal irradiance (DNI; $W/m^2$), global horizontal irradiance (GHI; $W/m^2$), temperature (°C) and wind speed (m/s). We obtained this dataset for Paris and Ile-de-France (49°N, 002.5°E) for 2019 with a tool SIREN, which calculates these parameters from climate reanalysis data MERRA-2 [27]. However, a direct application of this data can lead to an overestimate of the PV electricity production [9]. Thus, DHI, DNI, and GHI were reduced with a coefficient of 0.8, which was inferred from the analyses for Japanese cities [16]. The calculated capacity factor of 11.1% by SAM was a good agreement with the observed value of 11.0% in 2019 for Paris. The year 2019 was the one with the highest capacity factor in the period 2014-2020 (minimum 9.6% and mean 10.3%). In SAM, we chose the tilt angle of the panel to be 40° and its azimuth to be 180° for Paris and Ile-de-France.

2.5. Electricity demand and cost

The hourly electricity demand for Ile-de-France is available for the year 2019 by "Réseau de Transport d'Electricité" [28](Fig. 2). For 2030, we rescaled the 2019 demand by a factor 1.08 to match the electricity demand of EV for the "PV + EV" scenarios. The increased electricity demand was calculated under an assumed current EV efficiency of 17.2 kWh/100km. This efficiency corresponds to the mean use (mean between winter and summer conditions) of a Renault Zoé in city driving conditions. The driving patterns were assumed to be the same in 2019 and 2030. On the other hand, data for hourly electricity demand were not available for Paris. However, data for annual demand are given by Enedis, the power grid operator for the whole city [29]. For Paris, we thus used the hourly electricity demand of Ile-de-France multiplied by a factor of 0.18 based on the annual demand data of two regions. For 2030, we calculated that the demand increase due to EV charge is 2%, from deriving distance and EV efficiency resulting in the same scaling factor of 0.18 for Paris.

Concerning the prices of electricity, all our analyses for Paris and Ile-de-France used a weighted average price of 0.16 €/kW from the prices for household (0.178 €/kW) and industry (0.105 €/kW) [30] with annual electricity consumptions of 22.7 TWh and 7.3 TWh in 2019 [31], respectively. When we considered FIT, we used 0.04 €/kWh for both regions, which was an average day-ahead market price of electricity in 2019 for France [32]. We used the same price of electricity in 2019 and 2030. As the price of electricity has been rising in France for the past 30 years [30], particularly strongly in the recent past in the midst of geopolitical turmoil, it is likely to continue rising in the foreseeable future, which adds additional values on decentralized energy systems [17].

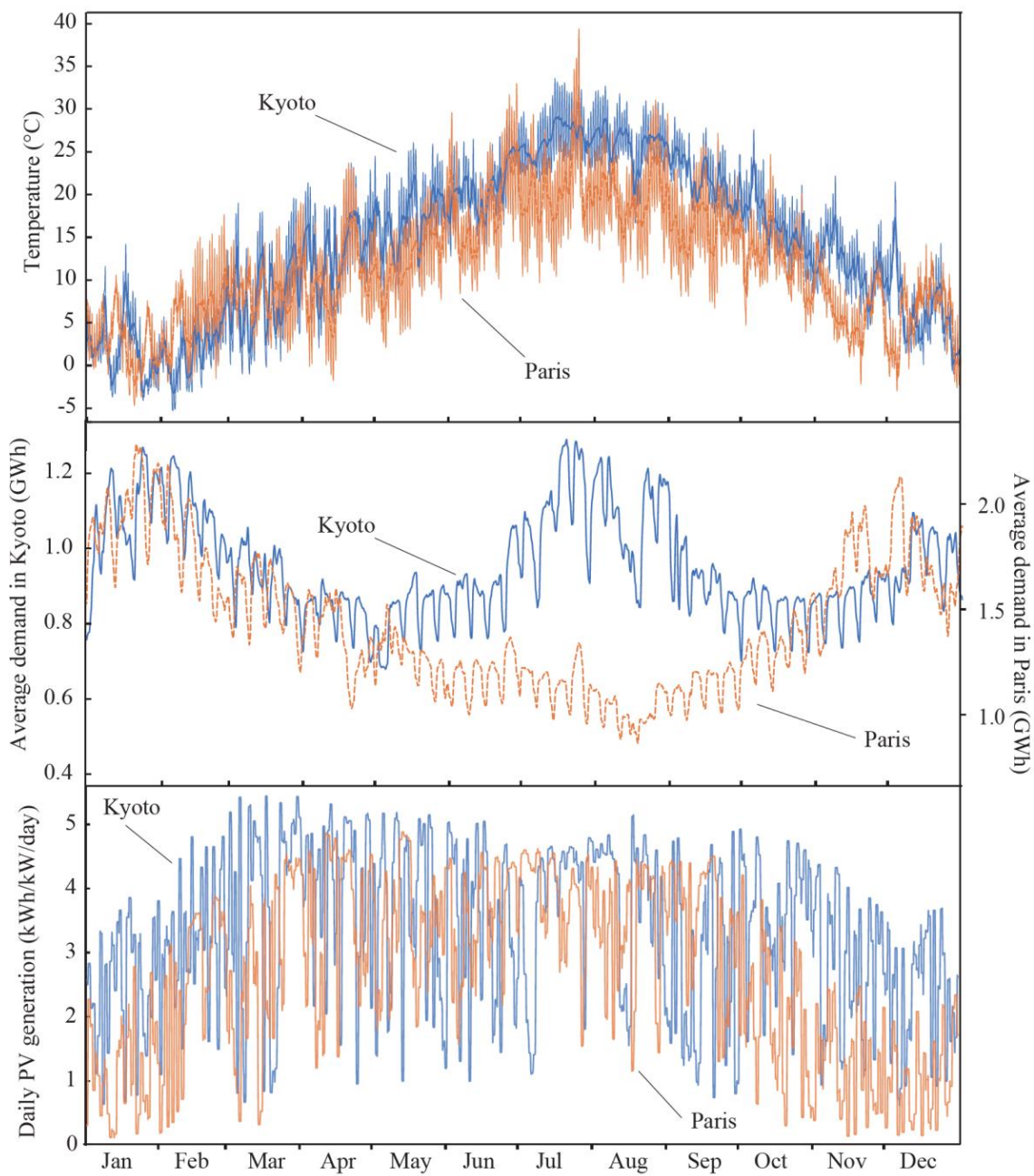

Figure 2. Temperature, average daily electricity demand, and daily PV generation for Paris in 2019 and Kyoto in 2018. The data for Kyoto are from [22].

2.6. PVs

The roof areas of Paris and Ile-de-France are 31 and 402 km$^2$, respectively [33]. To convert the roof area into the maximum capacity of PV, we used an estimate of 7 m$^2$ for 1 kWh of PV for both 2019 and 2030, which corresponds to the use of about 70% of the total rooftop area of the city [16]. In 2019, the cost of a PV-system depends on the surface area of panels for each roof. We used the 5-10 m$^2$ price (Table 3), which is higher than the price of 10-100 kW (thus 50-500 m$^2$) based on the dataset given by the Apur [34]. Using the higher price, our results on energy cost saving by PV system should be considered conservative. Maintenance prices were given as 22.5 €/kW/yr [34] and assumed to be the same in 2030. BloombergNEF (BNEF) estimates the global price for fixed-axis utility scale PV systems in 2030 to be 69% of 2020 prices (all costs included) [35]. We applied the same reduction to our residential prices, which leads to the prices for 2030 given in Table 3. A conversion rate of 1.12$/€ was used (2019 mean).

Table 3. Prices of PVs assumed in this study.

|  | Paris | Ile-de-France |
|---|---|---|
| PV, 2019, installation and inverter included (€/W) | 1.9 | 1.9 |
| PV, 2030, installation and inverter included (€/W) | 1.31 | 1.31 |

2.7. EVs and vehicle statistics

We assumed that EVs are equipped with a 40kWh battery [16]. We set the power charge to 6 kW, and discharge can be expected with the same power for a V2H system [16]. It was projected that Europe would reach price parity between EV (including home chargers) and ICE during the period 2025-2030 [36]. The additional costs of having EV plus V2H system is estimated to be around 25 €/kWh in 2030 [16]. The cost of battery replacement of 91 € per battery (when the capacity is reduced to 80% of the initial capacity) is included in EV price [16]. We limit the use of EV batteries only in the range from 50 to 95% of their charge to prevent battery degradations and to allow EV owners to use their EVs for short trips anytime.

Number of cars and annual driving distance are given in Table 2. We assumed that the number of cars, the proportion of cars used during weekdays, and the average distance of moving cars in Paris and Ile-de-France would be the same between 2019 and 2030. It should be noted that while Paris and Ile-de-France replace ICEs with EVs, the number of vehicles may be reduced. The city of Paris is promoting the use of bicycles instead of cars, which is also an important decarbonization measure [37]. A decrease in the use of vehicles, therefore an increase of the parked time of vehicles, might raise the potential of "PV + EV". However, a significant change in the number of vehicles would reduce the total battery capacity and thus could affect the results of the "PV + EV" scenario.

# 3. Results

3.1. Paris vs. Ile-de-France

For Paris, "PV only" scenario can bring benefits already in 2019, and the benefits can further increase in 2030 (Fig. 3a). There are optimum PV capacities of 2.7 GW and 3.6 GW with FITs for 2019 and 2030, respectively (Fig. 3a; Table 4). Figure 3b shows that in the "PV only" case, surplus PV electricity start increasing from around 40%. From this point, FIT starts to play an important role to increase NPVs and optimal PV capacity. In addition, above this threshold, self-sufficiency diverts from energy sufficiency, and self-consumption starts to decline (Fig. 3b). Concerning "PV+EV", NPVs become larger than zero above 10% coverage in 2030, and NPVs continue rising above 70% (Fig 3a). Thus, the rooftop PV economics reaches the maximum with the maximum usage of the rooftop area (70%) in Paris by coupling with EVs. Self-consumption is 100% even at the maximum rooftop usage (71%) for "PV + EV" system with self-sufficiency and energy sufficiency being equal (Fig. 3c).

For Ile-de-France, there are optimums for PV capacity in all the scenarios (Fig. 3d). For "PV only", FITs make a difference with a roof coverage of above around 20% (Fig. 3d), which is much smaller than those for Paris owing to larger rooftop area per capita (Fig. 3a). When the rooftop coverage of PV becomes larger than around 20%, PV electricity production becomes greater than the demand in many hours of the year, resulting in a lower self-consumption and a larger difference between self-sufficiency and energy sufficiency (Fig 3e), which occurs in smaller roof area coverage than that of Paris. In addition, NPVs reach peaks in 54% and 60 % rooftop coverage, and abruptly drops, which is the result of EV battery replacement as a large investment for the replacement reduces the optimal PV capacity (Fig 3a). It is noted that energy sufficiency in Ile-de-France reaches 78%, which is much larger than 34% of Paris. Concerning "PV + EV", NPVs become larger than zero above 8-10% coverage in 2030. In contrast to Paris, self-consumption exhibits a slight decrease from 62% of the roof coverage with increasing surplus electricity. Accordingly, self-sufficiency is lower than energy sufficiency above 62 % of the rooftop coverage (Fig. 3f).

The results above indicates that Paris, where energy demand per area is high and the rooftop area is relatively limited, is better suited to have "PV only" below 50% of the rooftop coverage. For Ile-de-France with a relatively large rooftop area, on the other hand, PV-generated electricity can be quickly saturated with a small roof coverage (Fig. 3e). In this instance, PV + EV scenario has more advantages than PV only scenario, as long as the systems will be implemented at a substantial scale (i.e., above the threshold rooftop area). If PV only scenario proceeds further in Ile-de-France, it could lead to an overproduction of electricity, requiring curtailment with reduction in PV economics, as is already happening in some areas with a large amount of renewables. It is noted that NPVs in Figure 3 do not include cost saving from elimination of gasoline/dieses expenses. Thus, rooftop PV + EV owners have larger cost benefits than the results shown in Figure 3.

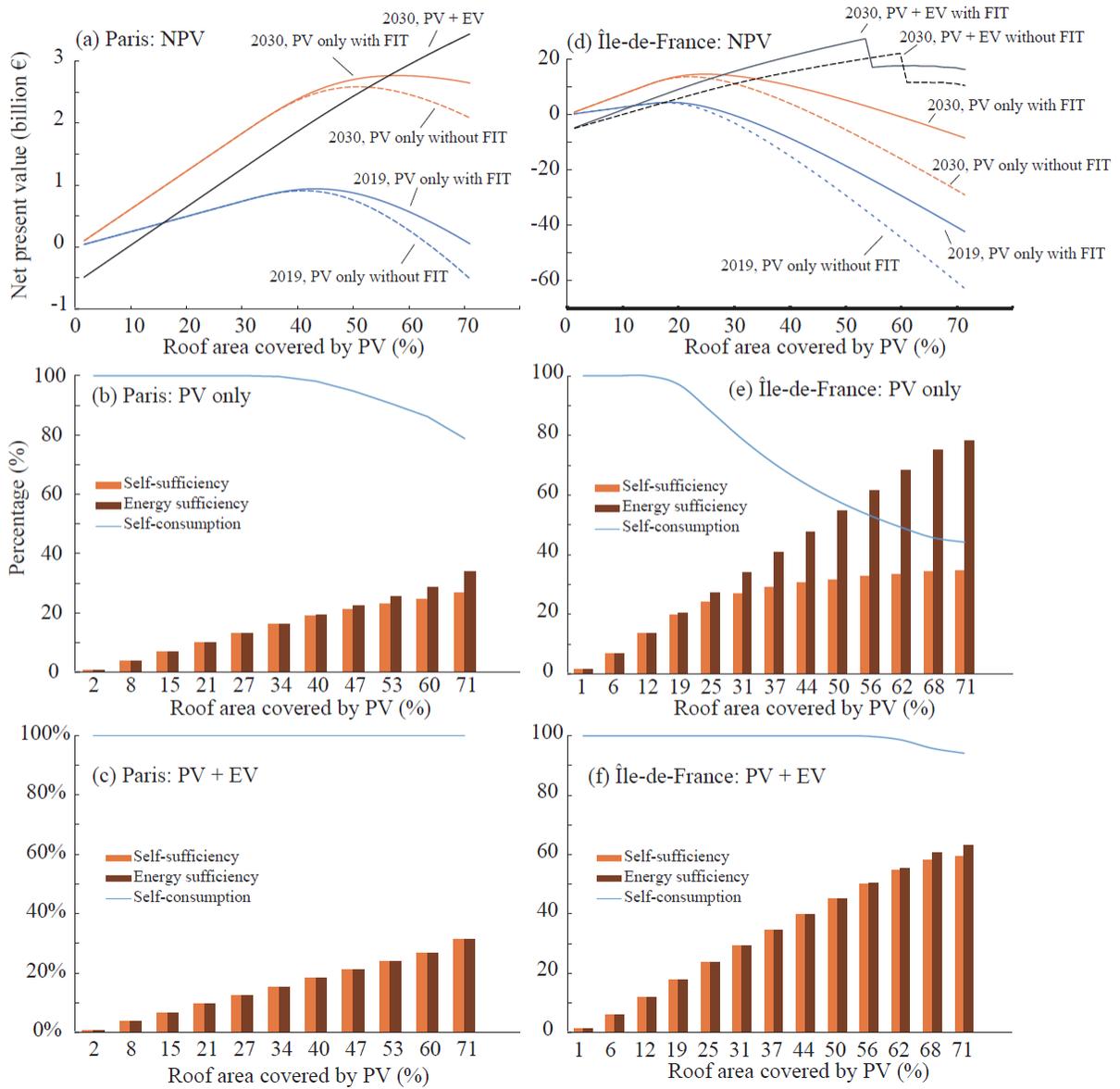

Figure 3. "PV only" and "PV + EV" potentials for Paris (left panels) and Ile-de-France (right panels). Note that NPVs do not include cost savings from gasoline/diesel expenens.

Table 4. Decarbonization indicators for Paris and Ile-de-France in 2019 and 2030. Cost-savings include those from gasoline/diesel expenses as in [22]. The percentages indicated for the optimal PV capacity are the corresponding rooftop area in the city.

| Paris | 2019 "PV only" | | 2030 "PV only" | | 2030 "PV+EV" | |
|---|---|---|---|---|---|---|
| | with FIT | without FIT | with FIT | without FIT | with FIT | without FIT |
| Optimal PV capacity (GW) | 2.7 (44%) | 2.5 (44%) | 3.6 (58%) | 3.2 (52%) | 4.4 (71%) | 4.4 (71%) |
| Self-consumption (%) | 97 | 98 | 87 | 92 | 100 | 100 |
| Self-sufficiency (%) | 20 | 19 | 24 | 23 | 31 | 31 |
| Energy sufficiency (%) | 21 | 19 | 28 | 25 | 31 | 31 |
| Cost saving (%) | 2 | 2 | 5 | 5 | 12 | 12 |
| $CO_2$ emission reduction (%) | 18 | 17 | 21 | 20 | 51 | 51 |

| Ile-de-France | 2019 "PV only" | | 2030 "PV only" | | 2030 "PV+EV" | |
|---|---|---|---|---|---|---|
| | with FIT | without FIT | with FIT | without FIT | with FIT | without FIT |
| Optimal PV capacity (GW) | 15 (19%) | 14 (17%) | 20 (25%) | 18 (22%) | 48 (54%) | 48 (60%) |
| Self-consumption (%) | 97 | 98 | 88 | 92 | 99 | 100 |
| Self-sufficiency (%) | 20 | 19 | 24 | 23 | 53 | 50 |
| Energy sufficiency (%) | 21 | 19 | 27 | 25 | 53 | 50 |
| Cost saving (%) | 1 | 1 | 4 | 4 | 21 | 22 |
| $CO_2$ emission reduction (%) | 13 | 12 | 14 | 14 | 79 | 77 |

3.2. Paris vs. Kyoto

The potential of the SolarEV City Concept differs not only by the geographical extent as exemplified by the comparison between Paris and Ile-de-France but also climate, latitude, electricity tariff, etc. Therefore, it is useful to compare Paris with Kyoto, Japan, where the SolarEV City Concept was first introduced [9]. There is a large difference in latitude between Paris and Kyoto. As Earth is round and the rotating axis is tilted, higher latitudes experience annually a lesser amount of solar irradiance and larger seasonal change of solar irradiance, which affects energy yields of PV. The slope and azimuth of PV panels also affect energy yields differently in different latitudes (Fig. 4). Flat PV panels in Paris produces about 24 % less energy in a year than that of Kyoto (Fig. 4). Maximum energy yields can be obtained in a 30°-50° slope in Paris and in a 30° slope in Kyoto with south faced azimuth (180°) (Fig. 4), where Paris has still 21% less energy yield than that of Kyoto. However, energy yields on southern faced façades (a slope of 90%) in Paris are only 9 % less than that of Kyoto (Fig. 4), indicating that façade PVs are relatively good options in higher latitudes.

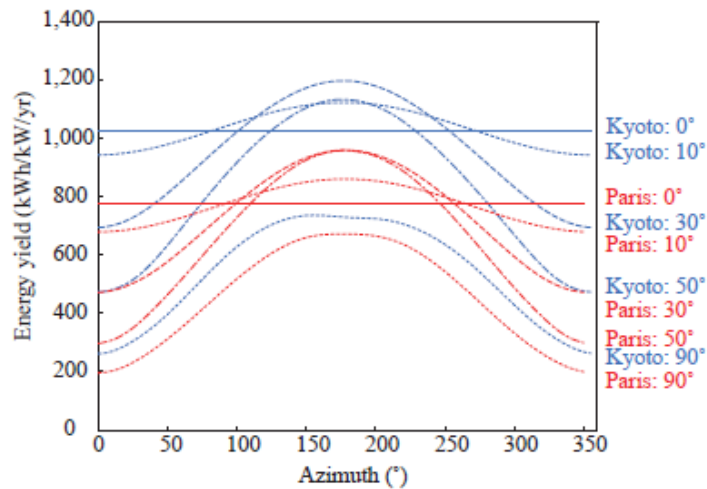

Figure 4. Energy yields with different slopes of rooftop PVs with various azimuths. Red lines represent Paris, and blue lines represent Kyoto.

On seasonal variation of PV energy yields, Paris has notable decline in electricity generation in winter compared to Kyoto (Fig. 5). The variation can be alleviated by having some slopes, but the winter decline in Paris is rather large, considering large electricity demands for heating. It is noted that wind power generates more electricity in winter in Europe [38]. That can compensate the solar power decline in the region.

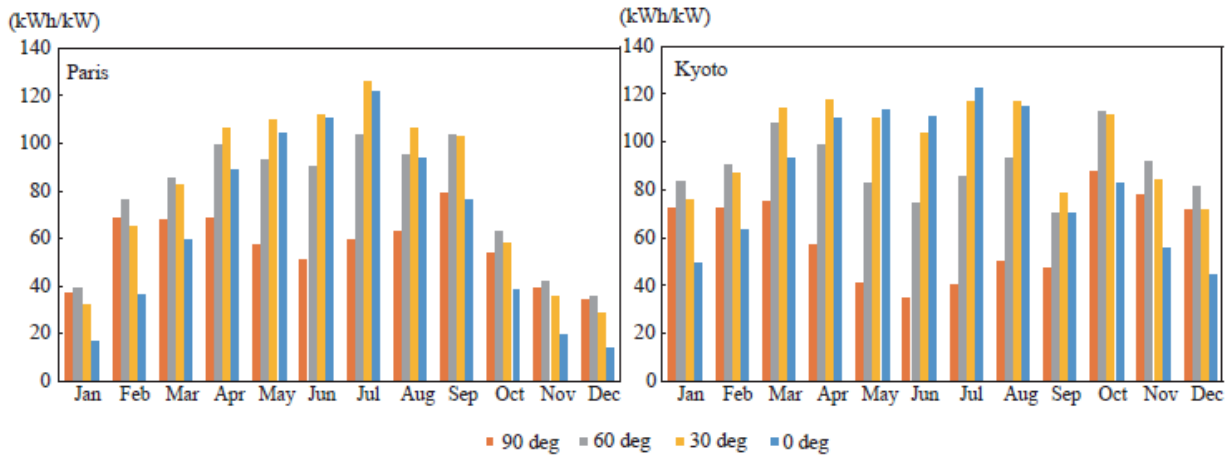

Figure 5. Monthly PV generation with various PV slopes in Paris and Kyoto.

Figure 6 illustrates average hourly electricity demand variations in a day for different months for Paris and Kyoto [22]. In Paris, hourly demand variation shows some consistent characteristic through the year with a first peak around noon then another peak around 19:00. During summer, demands are smallest and in winter demand reaches peaks (Fig. 2 and 6). On the other hand, daily demand profiles in summer and winter are different in Kyoto (Fig. 6). The largest electricity demand comes in summer with one large peak in the afternoon. In winter, two peaks develop, a morning peak around 9 pm and an evening peak around 7 pm. It is interesting to note that a drop in the lunch break only develops in Kyoto not in Paris (Fig. 6). Figure 6 also shows how electricity is supplied from the "PV + EV" system for both cities. In Paris, owing to smaller rooftop area per capita and less PV generation, the "PV + EV" system can only supply 31% of its demand. On the other hand, the "PV + EV" system in Kyoto can supply 76% of electricity demand, which includes the demand from EVs (Table 4 and 5).

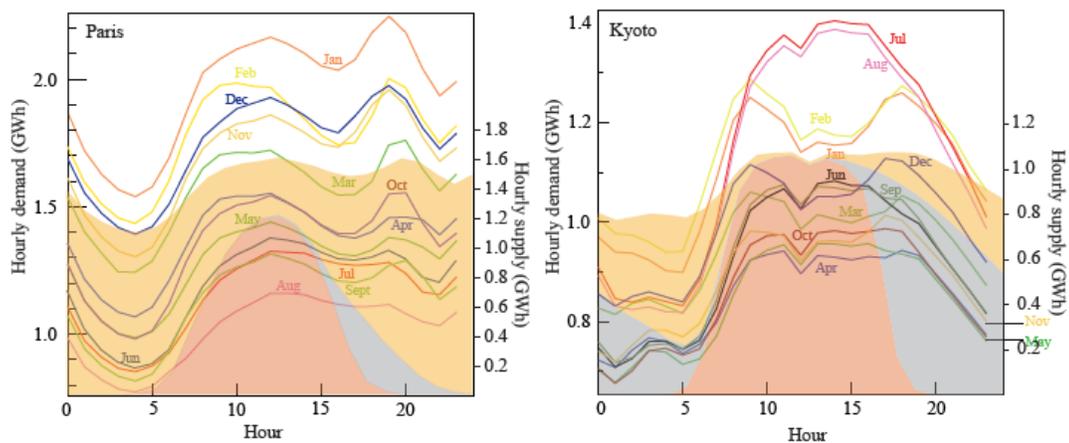

Figure 6. Hourly demand in a day for different months and annual-mean supply by "PV + EV" in a day. Electricity supplies from grid, PV, and battery are shown in orange, red, and grey, respectively for the "PV + EV" scenario with FIT in 2030 for both cities.

Table 5. Decarbonization indicators for Kyoto. Cost-savings include those from gasoline/diesel expenses as in [22]. The percentages indicated for the optimal PV capacity are the corresponding rooftop area in the city.

|  | 2019 "PV only" | | 2030 "PV only" | | 2030 "PV+EV" | |
| --- | --- | --- | --- | --- | --- | --- |
| **Kyoto** | with FIT | without FIT | with FIT | without FIT | with FIT | without FIT |
| Optimal PV capacity (GW) | 2.3 (22%) | 1.8 (17%) | 7.4 (72%) | 2.8 (27%) | 7.4 (72%) | 5.9 (57%) |
| Self-consumption (%) | 88 | 96 | 40 | 80 | 84 | 94 |
| Self-sufficiency (%) | 29 | 25 | 43 | 32 | 76 | 68 |
| Energy sufficiency (%) | 33 | 26 | 106 | 40 | 90 | 73 |
| Cost saving (%) | 4 | 4 | 18 | 11 | 31 | 29 |
| $CO_2$ emission reduction (%) | 27 | 23 | 40 | 30 | 78 | 70 |

Figure 7 shows cross-spectral coherence [39] between electricity demand vs. temperature and PV generation. Cross-sectional coherence for temperature vs. demand shows positive coherence (or correlation) in summer and negative coherence (correlation) in winter for periods longer than 64 hours for both Paris and Kyoto (Fig. 7). Paris shows wider area of significant correlation both in winter and summer for the periods of 64-512 hours and 1024-2048 hours than those of Kyoto (Fig. 7). For the plot of demand vs. PV generation, cross-spectral coherence become less significant except daily periods. Positive coherence becomes significant in Kyoto in summer for the periods of 512 hours, indicating that space cooling demands are coinciding with higher PV generation (Fig. 7).

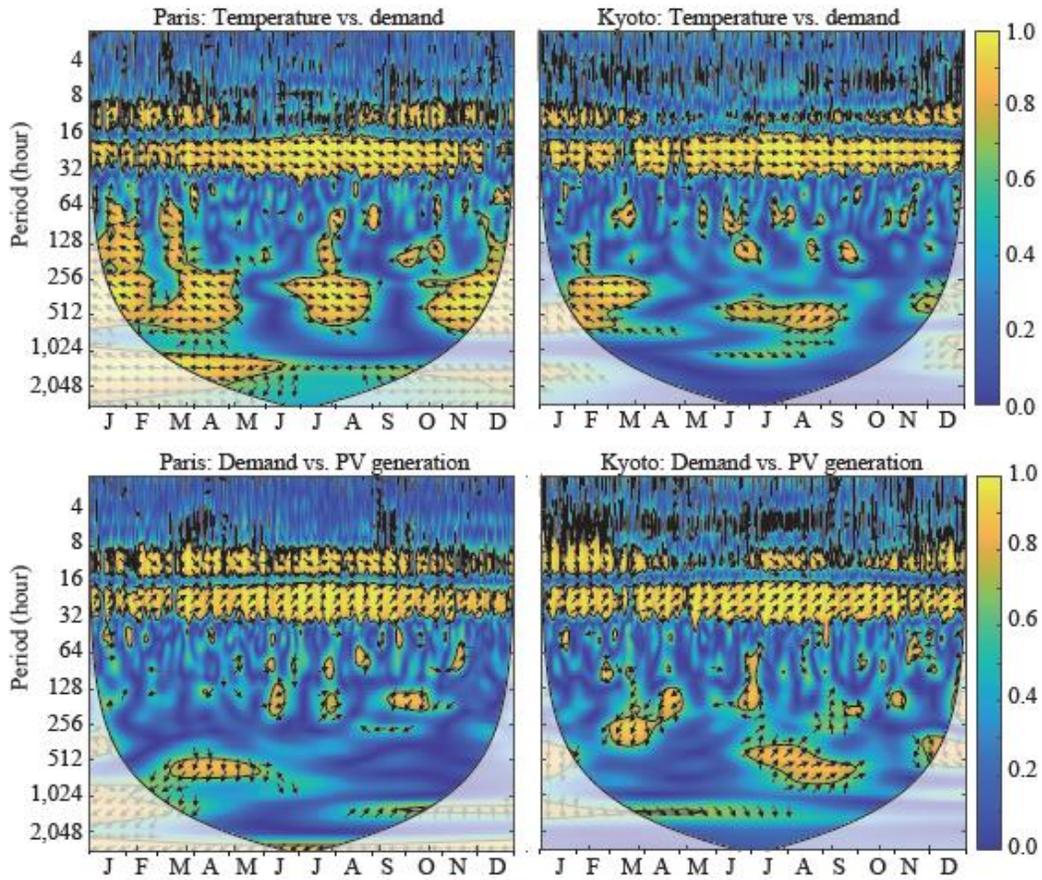

Figure 7. Cross-spectral coherence [39] for demand vs. temperature and PV generation. Allows pointing right indicate in-phase relationships, and vice versa. The areas enclosed with lines are significant in a 95% confidence. X-axis represents time, and y-axis represents periods.

Figure 8 shows Kyoto's decarbonization potentials for "PV only" and "PV + EV" in 2018 and 2030 as for Paris and Ile-de-France in Fig. 3. Kyoto's profiles are more similar to the those of Ile-de-France than that of Paris in terms of higher self-sufficiency and surplus electricity (Fig. 3 and 8). In Kyoto, the "PV + EV" system can provide 76 % (self-sufficiency) of electricity demand including the demands from EVs. In comparison with "PV only" system, coupling PV systems with EV battery can help increase self-consumption from 40 % to 84 % (Fig. 8 and Table 4).

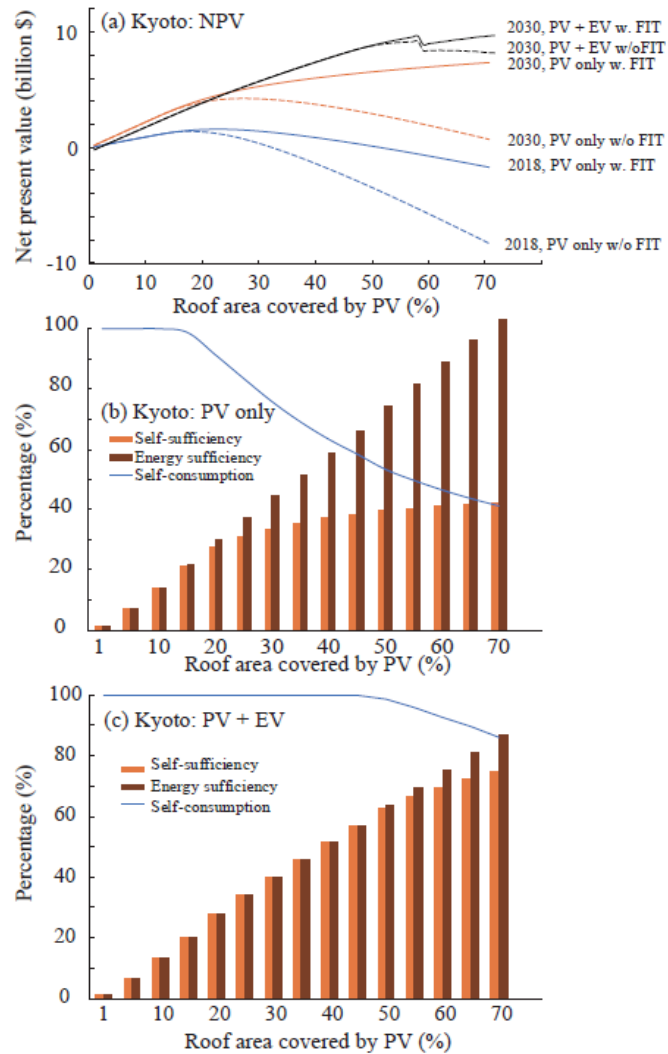

Figure 8. "PV only" and "PV + EV" potentials for Kyoto. "w." and "w/o" indicate with and without, respectively.

The differences between Paris and Kyoto on the effects of "PV + EV" originate from energy structure of the cities and climate. To understand the influence of climate on the "PV + EV" system for Paris, we apply the energy demand, the number of vehicles per capita (also annual driving distance), and rooftop area per capita of Kyoto to Paris as a sensitivity analysis, such that the energy structure becomes comparable with that of Kyoto in the analysis. Thus, the difference between two cities become only climate [26]. With these settings, self-consumption, self-sufficiency, and energy sufficiency of the "PV + EV" system for Paris is 87%, 62%, and 71% for the "PV + EV" system with the full usage (71 %) of rooftop area in the city. The self-sufficiency in Kyoto is 76 % (Table 5), which is higher by 14 points than that of Paris. Our sensitivity analysis indicates that the effects of the "PV + EV" system in Paris are lower than those for the city of Kyoto owing to the climate (e.g., higher latitude).

3.3. $CO_2$ emission reductions

France is already quite low carbon in terms of electricity generation owing to accumulated investments on nuclear power plants. In 2019, grid $CO_2$ emission factor for France was 0.063 $kg_{CO2}$/kWh [40], which can be compared with 0.352 $kg_{CO2}$/kWh in 2019 for Kyoto [41]. This indicates that the "PV + EV" system can reduce $CO_2$ emission from electricity consumption by 0.020 $kg_{CO2}$/kWh (= 0.063*0.31, where 0.31 is self-sufficiency for PV + EV) for the Paris's power system and 0.270 $kg_{CO2}$/kWh (= 0.352*0.76, where 0.76 is self-sufficiency for PV + EV) in 2019 for the Kyoto's. Therefore, the "PV + EV" system in Kyoto, where the current electricity generation has high $CO_2$ emission from coal- and gas-fired power plants, is 13.5 times more effective to lower the mean $CO_2$ content (/kWh) of power generation than that in Paris.

## 4. Discussion

Our model calculations showed that developing rooftop PV could bring economic benefits for both Paris and Ile-de-France. It was estimated to be the case already in 2019 and the benefits could increase in 2030 owing to declining costs. Up to a certain roof coverage, 50-60% of the total roof area for Paris and 20-30% for Ile-de-France, the production does not exceed the demand, resulting in a nearly 100% self-consumption at the macroscopic level. Above such threshold, FIT was shown to play a role. EVs used batteries are another option, which can be more profitable thanks to the combination of EVs with PV through V2H or V2B systems developed at the city or region level. V2H or V2B systems can add value to PV systems, allowing a high self-consumption and self-sufficiency, when the roof coverage exceeds the threshold coverage for the region. The systems also allow bypassing the classical battery storage, which is too expensive to be profitable. Finally, because the electricity production is low carbon in France due to the reliance on nuclear power (70.5% of generated electricity in 2019) [42], solar panels alone do not allow substantial $CO_2$ emission abatements. But the "PV+EV" scenario facilitates further $CO_2$ emission abatements directly or indirectly from the use of EV relative to that of ICE.

Technical difficulties may hinder the implementation of the PV systems on complicated roof surface and V2H and V2B systems in densely populated districts. Grid constraints could also be limiting factors. These factors may work against a widespread use of PV and PV+EV. Furthermore, there are also practical constraints for the implementation of "PV only" and "PV + EV" systems. Wind turbines are not well accepted in some French regions, and solar panels may not be socially accepted in a historic city such as Paris [43]. Most buildings in Paris are under co-ownership that would require an agreement on the PV installation and use, as well as associated cost and benefit sharing. Legal difficulties could also arise, since buildings situated less than 500 m away from a historical monument or landmark must ask for a specific permit for a PV installation. The present roof coverage of PVs is less than 0.3% in Paris today. While the use of EV is becoming more widespread in Paris these days, not all buildings have sufficient parking space for EVs. Such infrastructure constrain may hinder the implementation of V2H systems in Paris. However, such barriers may be lower in Ile-de-France than in Paris. Even if the systems bring net benefits at some point of the long-term project period, the requirement of a large head investment could also be an economic barrier. To address those issues, strong political decisions, accompanied by legal and financial support and stakeholder engagement, may be needed.

## 5. Conclusions

While the SolarEV City Concept can face certain implementation barriers in Paris and Ile-de-France and may only lead to a modest reduction in GHG emissions in France, it can improve overall energy efficiencies by making use of PVs supplemented by the storage capacity of EV and may facilitate a low-carbon shift in transport as part of urban transformation. The SolarEV City Concept can be one of the pillars for transformation toward sustainability and jointly addressed with other pillars. Our analysis suggested that Ile-de-France would be a more promising venue for deploying the SolarEV City approach than Paris, or the deployment should be done in all the Ile-de-France administrative region including Paris or jointly for Paris and Ile-de-France such that large surplus electricity in suburb can be consumed in the center of the city. At the regional level, the storage brought by EVs can be useful with a relatively low rooftop PV coverage in Ile-de-France (20-30%). Ile-de-France is less susceptible to implementation barriers than Paris. Synergies between PVs and EVs cannot be expected in Paris below the rooftop PV coverage of 50%-60% in our estimate. Our analyses indicate that the high-latitude locations of Paris and Ile-de-France is not an advantage for the "PV + EV" system, but it can be complemented with high wind power potential in winter in the region. Although low-carbon electricity is already realized partially in France by nuclear power, the SolarEV City Concept may help take one more step forward toward zero-carbon energy systems in Paris.

**Acknowledgement**

K.T. benefited from State assistance managed by the National Research Agency in France under the Programme d'Investissements d'Avenir under the reference ANR-19-MPGA-0008. The project is supported by JSPS KAKEN: 23K11520 to T.K.

**CRediT authorship contribution statement**

Paul Deroubaix: Methodology, Formal analysis, Investigation, Writing - Original Draft, Visualization. Takuro Kobashi: Conceptualization, Methodology, Validation, Formal analysis, Investigation, Writing - Original Draft, Writing - Review & Editing, Visualization, Project administration. Léna Gurriaran: Investigation, Writing - Original Draft. Fouzi Benkhelifa: Investigation, Writing - Review & Editing. Philippe Ciais: Investigation, Writing - Review & Editing, Funding acquisition, Project administration. Katsumasa Tanaka: Conceptualization, Investigation, Writing - Original Draft, Writing - Review & Editing, Supervision, Project administration.

**Declaration of Competing Interest**

The authors declare that they have no known competing financial interests or personal relationships that could have appeared to influence the work reported in this paper.

**Data availability**

Supplementary data is available at https://data.mendeley.com/datasets/t93bh6bthj/1.

**References**

[1] Tanaka K, O'Neill BC. The Paris Agreement zero-emissions goal is not always consistent with the 1.5 °C and 2 °C temperature targets. Nat Clim Chang 2018;8:319–24. https://doi.org/10.1038/s41558-018-0097-x.

[2] van Soest HL, den Elzen MGJ, van Vuuren DP. Net-zero emission targets for major emitting countries consistent with the Paris Agreement. Nat Commun 2021;12. https://doi.org/10.1038/s41467-021-22294-x.

[3] European Union. Regulation (EU) 2021/1119 of the European Parliament and of the Council of 30 June 2021 establishing the framework for achieving climate neutrality and amending Regulations (EC) No 401/2009 and (EU) 2018/1999 ('European Climate Law') n.d. https://eur-lex.europa.eu/legal-content/EN/TXT/?uri=CELEX:32021R1119 (accessed May 28, 2023).

[4] IPCC. Climate Change 2022: Mitigation of Climate Change. Contribution of Working Group III to the Sixth Assessment Report of the Intergovernmental Panel on Climate Change. Cambridge, United Kingdom: 2022.

[5] Linton S, Clarke A, Tozer L. Technical pathways to deep decarbonization in cities: Eight best practice case studies of transformational climate mitigation. Energy Res Soc Sci 2022;86. https://doi.org/10.1016/j.erss.2021.102422.

[6] IEA. Net zero by 2050: A roadmap for the global energy sector. 2021.

[7] IEA. World Energy Outlook 2022. 2022.

[8] Kobashi T, Jittrapirom P, Yoshida T, Hirano Y, Yamagata Y. SolarEV City concept: Building the next urban power and mobility systems. Environmental Research Letters 2021;16:024042. https://doi.org/10.1088/1748-9326/abd430.

[9] Kobashi T, Yoshida T, Yamagata Y, Naito K, Pfenninger S, Say K, et al. On the potential of "Photovoltaics + Electric vehicles" for deep decarbonization of Kyoto's power systems: Techno-economic-social considerations. Appl Energy 2020;275:115419. https://doi.org/10.1016/j.apenergy.2020.115419.

[10] Gumilang R, Welo U, Siagian R, Asmara B, Dyah S, Ichihara J, et al. Equitable , affordable , and deep decarbonization pathways for low-latitude developing cities by rooftop photovoltaics integrated with electric vehicles. Appl Energy 2023;332:120507. https://doi.org/10.1016/j.apenergy.2022.120507.

[11] Chang S, Cho J, Heo J, Kang J, Kobashi T. Energy infrastructure transitions with PV and EV combined systems using techno-economic analyses for decarbonization in cities ☆. Appl Energy 2022;319:119254. https://doi.org/10.1016/j.apenergy.2022.119254.

[12] Liu J, Li M, Xue L, Kobashi T. A framework to evaluate the energy-environment-economic impacts of developing rooftop photovoltaics integrated with electric vehicles at city level. Renew Energy 2022;200:647–57. https://doi.org/10.1016/j.renene.2022.10.011.

[13] Arowolo W, Perez Y. Rapid decarbonisation of Paris, Lyon and Marseille's power, transport and building sectors by coupling rooftop solar PV and electric vehicles. Energy for Sustainable Development 2023;74:196–214. https://doi.org/10.1016/j.esd.2023.04.002.
[14] Insee. Statstics on the France economy 2021. https://www.insee.fr/en/accueil (accessed April 21, 2023).
[15] Capital. Comment Paris se met à l'énergie solaire 2019. https://www.capital.fr/economie-politique/comment-paris-se-met-a-lenergie-solaire-1344227 (accessed April 21, 2023).
[16] Kobashi T, Jittrapirom P, Yoshida T, Hirano Y, Yamagata Y. SolarEV City concept: Building the next urban power and mobility systems. Environmental Research Letters 2021;16. https://doi.org/10.1088/1748-9326/abd430.
[17] Hoppmann J, Volland J, Schmidt TS, Hoffmann VH. The economic viability of battery storage for residential solar photovoltaic systems – A review and a simulation model. Renewable and Sustainable Energy Reviews 2014;39:1101–18.
[18] Das R, Wang Y, Putrus G, Kotter R, Marzband M, Herteleer B, et al. Multi-objective techno-economic-environmental optimisation of electric vehicle for energy services. Appl Energy 2020;257. https://doi.org/10.1016/j.apenergy.2019.113965.
[19] Han X, Garrison J, Hug G. Techno-economic analysis of PV-battery systems in Switzerland. Renewable and Sustainable Energy Reviews 2022;158. https://doi.org/10.1016/j.rser.2021.112028.
[20] Li Y, Liu C. Techno-economic analysis for constructing solar photovoltaic projects on building envelopes. Build Environ 2018;127:37–46. https://doi.org/10.1016/j.buildenv.2017.10.014.
[21] Chang S, Cho J, Heo J, Kang J, Kobashi T. Energy infrastructure transitions with PV and EV combined systems using techno-economic analyses for decarbonization in cities. Appl Energy 2022;319. https://doi.org/10.1016/j.apenergy.2022.119254.
[22] Kobashi T, Yoshida T, Yamagata Y, Naito K, Pfenninger S, Say K, et al. On the potential of "Photovoltaics + Electric vehicles" for deep decarbonization of Kyoto's power systems: Techno-economic-social considerations. Appl Energy 2020;275. https://doi.org/10.1016/j.apenergy.2020.115419.
[23] Blair N, Diorio N, Freeman J, Gilman P, Janzou S, Neises T, et al. System Advisor Model (SAM) General Description (Version 2017.9.5). 2018.
[24] Blair N, Diorio N, Freeman J, Gilman P, Janzou S, Neises TW, et al. System Advisor Model (SAM) General Description 2018.
[25] Kobashi T. Paris SAM files 2023. doi: 10.17632/t93bh6bthj.1 (accessed May 17, 2023).
[26] Dewi RG, Siagian UWR, Asmara B, Anggraini SD, Ichihara J, Kobashi T. Equitable, affordable, and deep decarbonization pathways for low-latitude developing cities by rooftop photovoltaics integrated with electric vehicles. Appl Energy 2023;332. https://doi.org/10.1016/j.apenergy.2022.120507.
[27] King A. SIREN: SEN's interactive renewable energy network tool. In: Sayigh A, editor. Transition towards 100% renewable energy, Cham: Springer; 2018, p. 536.
[28] RTE. éCO2mix - La consommation d'électricité en France 2023. https://www.rte-france.com/eco2mix/la-consommation-delectricite-en-france (accessed May 28, 2023).

[29]  Enedis. Open Data n.d. https://www.enedis.fr/open-data (accessed May 28, 2023).
[30]  Statistics report Energy Prices and Taxes for OECD Countries. 2020.
[31]  ROSE-ENERGIF. Réseau d'Observation Statistique de l'Énergie et des émissions de gaz à effet de serre en Île-de-France 2023. https://www.institutparisregion.fr/cartographies-interactives/energif-rose/ (accessed May 11, 2023).
[32]  RTE. Electricity report 2020 2023. https://bilan-electrique-2020.rte-france.com/market-prices-in-europe/?lang=en (accessed May 11, 2023).
[33]  apur. Geo data n.d. https://www.apur.org/fr/geo-data (accessed May 28, 2023).
[34]  ADEME. Etat du photovoltaïque en FRANCE 2019 2020. https://www.debatpublic.fr/sites/default/files/2021-07/HORIZEO-pvps-etat-photovoltaique-france-2019.pdf (accessed May 11, 2023).
[35]  BNEF. New Energy Outlook 2019. 2019.
[36]  Hitting the EV Inflection Point. 2021.
[37]  apur. Évolution des mobilités dans le Grand Paris Tendances historiques, évolutions en cours et émergentes 2021. https://www.apur.org/fr/nos-travaux/evolution-mobilites-grand-paris-tendances-historiques-evolutions-cours-emergentes (accessed April 21, 2023).
[38]  Heide D, von Bremen L, Greiner M, Hoffmann C, Speckmann M, Bofinger S. Seasonal optimal mix of wind and solar power in a future, highly renewable Europe. Renew Energy 2010;35:2483–9. https://doi.org/10.1016/j.renene.2010.03.012.
[39]  Grinsted A, Moore JC, Jevrejeva S. Application of the cross wavelet transform and wavelet coherence to geophysical time series. Nonlinear Process Geophys 2004;11:561–6. https://doi.org/10.5194/npg-11-561-2004.
[40]  European Environment Agency. Greenhouse gas emission intensity of electricity generation in Europe 2022.
[41]  KEPCO. Changes in grid emission factor for KEPCO 2019:1. https://www.kepco.co.jp/corporate/pr/2019/pdf/0730_2j_01.pdf (accessed March 10, 2020).
[42]  IEA. France 2021 Energy Policy Review 2021:1–211.
[43]  Liberation. A Paris, une toiture sous contrainte 2018. https://www.liberation.fr/france/2018/02/22/a-paris-une-toiture-sous-contrainte_1631704/ (accessed April 21, 2023).